\documentclass[11pt]{article}
\usepackage{amsmath,amssymb,amsfonts,amsthm,bm,bbm,cancel,wasysym}
\usepackage{epsfig,graphics,graphicx,epstopdf,caption,subcaption}
\usepackage[numbers,sort&compress]{natbib}
\graphicspath{{Charts/}}
\usepackage{array,booktabs,colortbl,colordvi,multirow}
\usepackage{colordvi,color,xcolor}
\usepackage{hyperref}
\usepackage{rotating}
\usepackage{comment}
\usepackage{feynmf}

\usepackage[symbol]{footmisc}
\renewcommand{\thefootnote}{\fnsymbol{footnote}}

\begin{document}

\begin{center}

{\Large {\bf Interpreting the Hubble tension with a cascade decaying dark matter sector}}\\

\vspace*{0.75cm}

{Quan Zhou, Zixuan Xu and Sibo Zheng\footnote{Corresponding author: sibozheng.zju@gmail.com}}

\vspace{0.5cm}
{School of Physics, Chongqing University, Chongqing 401331, China}
\end{center}
\vspace{.5cm}

\begin{abstract}
\noindent
Hubble tension can be alleviated by altering either early- or late-time $\Lambda$CDM.
With only one of these effects introduced,
early dark energy remains the only solution capable of reducing the tension to the $3\sigma$ level or below.
In this work,  we instead consider a modification of the dark matter sector that incorporates both the early- and late-time
 effects, with the goal of achieving the largest possible value of $H_0$ within this framework.
As a realization of these two-fold effects, we study a cascade decaying dark matter model.
By fitting the model to the latest datasets of Planck CMB+ DESI BAO+Pantheon (+SH0ES),
we find that a 68$\%$ CL value of $H_{0}=68.76\pm0.35 (69.05^{+0.31}_{-0.27})$ km s$^{-1}$ Mpc$^{-1}$ with $\Delta \rm{AIC}=+22.0(18.4)$, and larger value of $H_0$ can be obtained by adjusting parameter priors  but with a cost of significantly increased value of $\Delta \rm{AIC}$.
These findings revise the earlier results on the tension level in the literature.
For completeness, we show that the parameter regions favored by the cosmological datasets are compatible with complementary limits arising from the Big Bang Nucleosynthesis, neutrino flux, and structure formation.

\end{abstract}

\renewcommand{\thefootnote}{\arabic{footnote}}
\setcounter{footnote}{0}
\thispagestyle{empty}
\vfill
\newpage
\setcounter{page}{1}

\tableofcontents

\section{Introduction}
The Hubble parameter $H$ is a fundamental cosmological observable characterizing the expansion speed of the Universe. 
Within the framework of $\Lambda$CDM, 
there exists a $\sim 5.8\sigma$ tension between an indirect measurement from the Planck 2018 data \cite{Planck:2018vyg} of Cosmic Microwave Background (CMB) reporting a value of  $H_{0}=67.36 \pm 0.54$ km s$^{-1}$ Mpc$^{-1}$ at the 68$\%$ confidence level (CL), 
and a direct measurement from the local experiment SH0ES \cite{Riess:2019cxk,Riess:2020fzl,Riess:2021jrx,Murakami:2023xuy} giving $H_{0}=73.01\pm 0.92$ km s$^{-1}$Mpc$^{-1}$.
Given that both measurements are highly precise,
 it is unlikely to attribute this discrepancy to systematical uncertainties. 
Consequently, the so-called Hubble tension is widely regarded as an indication of new physics beyond the $\Lambda$CDM cosmology.

Looking for a clue to reduce the Hubble tension, one has to keep in mind the following angle precisely measured by Planck, 
\begin{eqnarray}{\label{theta}}
\theta_{s}=\frac{r_{s}(z_{*})}{D_{A}(z_{*})},
 \end{eqnarray}
where $r_{s}(z)=\int^{\infty}_{z}c_{s}dz'/H(z')$ is the sound horizon and $D_{A}(z)=\int^{z}_{0}dz'/H(z')$ is the angular distance,
with $z_{*}$ the redshift at recombination. 
A nearly fixed value of $\theta_s$ in eq.(\ref{theta}) implies that the solution to the Hubble tension can be either an early- or late-time one, corresponding to lowering $r_{s}(z_{*})$ and increasing $D_{A}(z_{*})$ respectively. 
For comprehensive reviews on these solutions, see e.g, \cite{DiValentino:2021izs, Schoneberg:2021qvd, Khalife:2023qbu},
showing that either a single early- or late-time effect is insufficient to reduce the Hubble tension to the acceptable $3\sigma$ level
except an early dark energy (EDE) \cite{Karwal:2016vyq,Mortsell:2018mfj,Poulin:2018cxd}.

Therefore, the Hubble tension seems to favor the presence of both early- and late-time effects, 
as illustrated by $m_{e}+\sum m_{\nu}$ \cite{Khalife:2023qbu} and $m_{e}+\Omega_{K}$ \cite{Khalife:2023qbu,Sekiguchi:2020teg}, 
which can reduce the tension to $\sim 3\sigma$ level.
Instead of ad hoc combinations of those two-fold effects, 
in this work we investigate a cascade decaying DM (CDDM) sector that incorporates the two-fold effects.
In the framework of CDDM model, the cold dark matter (CDM) of $\Lambda$CDM is replaced by a sector where a parent particle decays to produce relativistic DM in the early-time Universe, and the DM decays instead of being absolutely stable in the late-time Universe.

Previously, altering the CDM component to reduce the Hubble tension has been studied in the literature.  For the relativistic effect of non-CDM on $H_0$, 
refs. \cite{Blinov:2020uvz, deJesus:2022pux, daCosta:2023mow} have reported an optimistic value of $H_0 \geq 70$ km s$^{-1}$ Mpc$^{-1}$.
For the late-time decay of DM into (dark) photons,
refs.\cite{Vattis:2019efj,Pandey:2019plg,Clark:2020miy,FrancoAbellan:2021sxk,Simon:2022ftd, Davari:2022uwd, Anchordoqui:2022gmw,Alvi:2022aam,McCarthy:2022gok,Xu:2023hkc} 
have shown that Planck data allow a relatively milder increase of $H_0$.\footnote{Late-time DM decaying into other SM particles has been used to explain the XENON1T anomaly \cite{Xu:2020qsy} and the neutrino event \cite{Kohri:2025bsn}.} 
Specifically, we study the CDDM model different from these earlier studies with two main features:
\begin{itemize}
\item While contributing to the effective neutrino number,
the DM energy density evolves non-trivially with redshift before the DM fluid becomes non-relativistic;
\item The DM decays to neutrinos rather than the (dark) photons, which allows relatively smaller DM lifetime to trigger larger late-time effects.
\end{itemize}
These properties allow the CDDM model to achieve the largest possible value of $H_0$ within scenarios modifying the dark matter sector.

The rest of this paper is structured as follows.
Sec.\ref{model} discusses how to parametrize the CDDM,
where the relevant background and perturbation equations prepared for later numerical analysis are presented.
Sec.\ref{MCMC} carries out a Markov Chain Monte Carlo (MCMC) analysis of the CDDM model to the latest datasets including the DESI BAO, 
showing the dependences of numerical results especially $H_0$ on the \textbf{local $H_0$ prior} and parameter priors.
Sec.\ref{cons} is devoted to the Big Bang Nucleosynthesis (BBN) limits from the early Universe, neutrino flux and structure formation constraint from the late Universe, respectively,
illustrating that the parameter regions favored by the cosmological datasets are compatible with these constraints.
Finally, we conclude in Sec.\ref{con}.

\section{Cascade decaying dark matter sector: parametrization}
\label{model}
In this work we consider the CDDM composed of two species $\chi_m$, $\chi_M$ with mass $m$ and $M$ respectively.  
The $\chi_M$ particle, which is unstable with lifetime $\tau_M$, decays to produce $\chi_m$ in the early Universe as
\begin{equation}\label{decay1}
\chi_{M} \rightarrow \chi_{m}+ X, 
\end{equation}
where $X$ denotes a SM final state. This decay induced $\chi_m$ particles are initially relativistic, 
but later become non-relativistic prior to the matter dominated epoch due to cosmic expansion.
Afterward, the $\chi_m$ particles, serving as a cold DM with decay width $\Gamma_{m}$, decay into the SM neutrinos in late-time Universe as 
\begin{equation}\label{decay2}
\chi_{m} \rightarrow \nu+\bar{\nu},~~(\Gamma^{-1}_{m}\geq 100~\rm{Gyr}).
\end{equation}
In the rest of this section, we will show how to parametrize the CDDM model, collect the set of independent model parameters, 
and derive background and perturbation equations for later numerical analysis.

\subsection{Production}
Regarding the early-time decay in eq.(\ref{decay1}),  the input parameters are composed of 
\begin{equation}\label{input1}
\{\rho_{M,0}, \tau_{M}, M, m, m_{X}\},
\end{equation}
where $\rho_{M,0}$ denotes the present-day $\chi_M$ energy density if it does not decay and $m_X$ the $X$ mass.
In order to produce the relativistic $\chi_m$ particles through this decay, 
one has to assume $M>>m$ and $M>>m_{X}$. 
In this situation the mass parameter $m_X$ in eq.(\ref{input1}) can be simply neglected. 
Moreover, as $\chi_m$ serves as the DM after the matter-dominated Universe, 
the present-day $\chi_m$  relic density has to accommodate the observed DM relic density if its does not decay in the late-time Universe,
which means that $\omega_{m,0}=\Omega_{m}h^{2}=(\rho_{m,0}/\rho_{\rm{c}})h^{2}$ is nearly fixed with $\rho_{c}$ the critical energy density and $\rho_{m,0}$ satisfying
\begin{equation}\label{equality}
\frac{\rho_{m,0}}{\rho_{M,0}}=\frac{m}{M}.
\end{equation}
Eq.(\ref{equality}) implies that $\rho_{M,0}$ can be fixed by adjusting the mass parameters $m$ and $M$.
As a result,  in eq.(\ref{input1}) we are left with the following three independent parameters
\begin{equation}\label{input1f}
\{\tau_{M}, M, m\}.
\end{equation}

To see how the parameters in eq.(\ref{input1f}) affect the value of $H_0$, we now derive the explicit forms of $\rho_{m}$ and $\rho_{M}$ as functions of time $t$.
First, for the non-relativistic $\chi_M$ with decay its energy density evolves as 
\begin{eqnarray}{\label{rho2}}
\frac{d\rho_{M}}{dt}+3H\rho_{M}=-\tau^{-1}_{M}\rho_{M},
 \end{eqnarray}
which gives 
\begin{eqnarray}{\label{rho2s}}
\rho_{M}(t)=\rho_{M,0}a^{-3}e^{-t/\tau_{M}}.
 \end{eqnarray}
 Second, the unperturbed Boltzmann equation for $\chi_m$ distribution function $f_m$ is given by \cite{Aoyama:2014tga}
\begin{eqnarray}{\label{df}}
\frac{\partial f_{m}}{\partial \tau}=\frac{a\rho_{M,0}e^{-t_{q}/\tau_{M}}}{4\pi\tau_{M} M\mathcal{H}q^{3}}\delta(\tau-\tau_{q})
 \end{eqnarray}
 where $\tau$ is the conformal time, $\mathcal{H}$ the conformal Hubble rate, $q=a(\tau_{q})p_{\rm{max}}$ the comoving momentum with $p_{\rm{max}}\approx M/2$ the decay induced momentum without suffering from the cosmic expansion.
 Given the distribution function the energy density reads as \cite{Ma:1995ey}
 \begin{eqnarray}{\label{rho1}}
\rho_{m}(t)=\frac{1}{a^{4}(t)}\int q^{2}dqd\Omega \epsilon f_{1}(q)
 \end{eqnarray}
 where $\epsilon=\sqrt{q^{2}+m^{2}a^{2}(t)}$ is the comoving energy. 
 Substituting the solution of eq.(\ref{df}) into eq.(\ref{rho1}) gives us the explicit form of $\rho_{m}(t)$.
 
Using eq.(\ref{equality}),  
we can rewrite  eq.(\ref{rho1}) as \cite{Blackadder:2014wpa}
\begin{equation}\label{rho1n}
\rho_{m}(t)\approx\frac{\rho_{m,0}\tau^{-1}_{M}}{a^{3}(t)}\int^{t}_{0}e^{-t_{D}/\tau_{M}}\sqrt{1+\frac{M^{2}}{4m^{2}}\left(\frac{a_{D}}{a(t)}\right)^{2}}dt_{D},
\end{equation}
where the subscript ``D" refers to decay. 
Compared to the time of radiation-matter equality with $t_{\rm{eq}}\approx1.6\times10^{12}$ sec,
if the decay takes place at time $t_D$ late enough to let the $a_{D}/a_{\rm{eq}}$-term dominate over unity, 
then one finds $\rho_{m}$ scales as $\sim a^{-4}$, suggesting that $\chi_m$ behaves as a relativistic particle.\footnote{This relativistic behavior can be also interpreted in terms of equation of state of $\chi_m$, see \cite{Blackadder:2014wpa} for details.}
Under this circumstance, it contributes to the effective neutrino number at the time of radiation-matter equality
\begin{eqnarray}\label{deltaN}
	\Delta N_{\rm{eff}}(t_{\rm{eq}})= \frac{\rho_{m}(t_{\rm{eq}}) }{\rho_{1\nu }(t_{\rm{eq}})}
	\approx \frac{\rho_{m,0}\tau^{-1}_{M} }{a^{3}(t_{\rm{eq}})\rho_{1\nu }(t_{\rm{eq}})} \frac{M}{2m} \int_{0}^{t_{\rm{eq}}}e^{-t_{D}/\tau_{M}}\sqrt{\frac{t_{D}}{t_{\rm{eq}}}}\,\,dt_{D}
\approx 1.19 \times \sqrt{\frac{\tau_{M}}{t_{\rm{eq}}}}\frac{M}{m},
\end{eqnarray}
where $\rho_{1\nu}$ is the energy density of a single neutrino species, 
and $a_{D}/a_{\rm{eq}}\approx (t_{D}/t_{\rm{eq}})^{1/2}$ during the radiation-dominated epoch has been used.
Eq.(\ref{deltaN}) shows that a large value of $H_0$ can be obtained by choosing a large $\tau_M$ and a large mass ratio of $M/m$.
For example, with  $\tau_{M}\sim 10^{4}$ sec and $M/m\sim 10^{3}$, we have $\Delta N_{\rm{eff}}(t_{\rm{eq}})\sim 0.1$. 

The early-time improvement on $H_0$ is however limited. 
First,  the Hubble rate enhanced at the early-time Universe leads to a smaller value of sound horizon $r_{s}(z_{*})$ with $z_{*}$ the redshift at recombination, which is subject to the CMB constraints as discussed in Sec.\ref{MCMC}. 
Second, the early-time decay induced energy injection into the SM thermal bath affects the light element abundances including D, $^{4}$He and $^{7}$Li,
which is therefore constrained by the BBN limits.
As discussed in Sec.\ref{BBN}, the BBN limits are stringent for $\tau_{M}> 10^{4}$ but become weak for smaller values of $\tau_{M}$. 
In this sense, we restrict to the parameter range of $\tau_{M}\leq 10^{4}$ in the following analysis.

\subsection{Decay}
Regarding the late-time decay in eq.(\ref{decay2}), the inputs are composed of 
\begin{equation}\label{input2}
\{\rm{Br}, \Gamma_{m}\},
\end{equation}
where $\rm{Br}$ is the branching ratio of this decay channel. 
If one simply chooses $\rm{Br}=1$ as we adopt here, 
$\Gamma_{m}$ is the only model parameter controlling the late-time deviations from $\Lambda$CDM. 

Due to the $\chi_{m}$ decay, the background equation of $\rho_{m}$ and $\rho_{\nu}$ is modified as \cite{Poulin:2016nat}
\begin{eqnarray}{\label{rho1late}}
\rho'_{m}+3\frac{a'}{a}\rho_{m}&=&-a\Gamma_{m}\rho_{m},  \nonumber\\
\rho'_{\nu}+4\frac{a'}{a}\rho_{\nu}&=& a \Gamma_{m}\rho_{m},  
 \end{eqnarray}
 respectively, where primes denote derivatives with respect to the conformal time.
 
 Moreover, the matter perturbation $\delta_{m}$ changes in synchronous gauge as
\begin{eqnarray}{\label{matter}}
\delta'_{m}=-\frac{h'}{2},
 \end{eqnarray}
with  $h$ one of the two scalar modes in this gauge.
The decay-induced changes in the perturbation equation of neutrino energy density $\delta_{\nu}$  in synchronous gauge evolve as \cite{Xu:2023hkc} 
\begin{eqnarray}{\label{radp1}}
\delta'_{\nu}+\frac{4}{3}\theta_{\nu}+\frac{2}{3}h'&=&a\Gamma_{m}\frac{\rho_{m}}{\rho_{\nu}}(\delta_{m}-\delta_{\nu}),  \nonumber\\
\theta'_{\nu}-\frac{k^{2}}{4}(\delta_{\nu}-4\sigma_{\nu})&=&-a\Gamma_{m}\frac{\rho_{m}}{\rho_{\nu}}\theta_{\nu},  \nonumber\\
\sigma'_{\nu}-\frac{4}{15}\theta_{\nu}-\frac{2}{15}h'-\frac{4}{5}\eta'+\frac{3}{10}kF_{\nu3}&=&-a\Gamma_{m}\frac{\rho_{m}}{\rho_{\nu}}\sigma_{\nu},\nonumber\\
F'_{\nu\ell}+\frac{k}{2\ell+1}[(\ell+1)F_{\nu\ell+1}-\ell F_{\nu\ell-1}]&=&0,~~~ \ell \geq 3,
\end{eqnarray}
by using \cite{Kaplinghat:1999xy}, where $k$ is the wavenumber and $F_{\nu\ell}$ are defined in \cite{Ma:1995ey, Poulin:2016nat}.

Similar to the early-time improvement, the late-time improvement on $H_0$ is also constrained.
On one hand, neutrinos due to the late-time decay of $\chi_m$ contribute to neutrino flux constrained by existing data as discussed in Sec.\ref{Neutrino}.
On the other hand, the late-time decay of $\chi_m$ results in a suppression (increase) on the matter power spectrum at small (large) scales \cite{Xu:2023hkc,Poulin:2016nat}, which is constrained by the existing data about matter power spectrum.

\section{MCMC analysis}
\label{MCMC}
Implementing the background and linear perturbation equations in Sec.\ref{model} into the Boltzmann solver \texttt{CLASS} \cite{Lesgourgues:2011re,Blas:2011rf} with a couple of modified branches therein, we now use \texttt{Cobaya} \cite{Torrado:2020dgo,cobaya2019} to carry out the MCMC analysis on the CDDM model.

\subsection{Datasets}
\label{data}
We fit the CDDM model in Sec.\ref{model} to the following datasets.
\begin{itemize}
\item  \textbf{CMB}: we use the Planck 2018 low-$\ell$ temperature and polarization likelihood \cite{Planck:2019nip}, the high-$\ell$ TT, TE, and EE power spectra from \cite{Rosenberg:2022sdy},
and the Planck 2018 lensing potential power spectrum \cite{Planck:2018lbu}. 
\item \textbf{BAO}: we adopt the BAO measurements from the DESI DR2 data release \cite{DESI:2025zgx}, which provide both isotropic and anisotropic distance constraints over the redshift interval covered by the survey, thereby offering a precise probe of the late-time expansion rate and distance-redshift relation.
\item \textbf{Type Ia supernovae}: we include Type Ia supernova distance measurements from the Pantheon+ \cite{Brout:2022vxf,Scolnic:2021amr} compilation.
\item \textbf{Local $H_0$ prior}: finally we include a local prior on the Hubble constant $H_0$ = 73.04 $\pm$ 1.04 km/s/Mpc from the SH0ES collaboration \cite{Riess:2021jrx}.
\end {itemize}
We clarify that the LSS data \cite{Heymans:2013fya,Planck:2013lkt} is not taken into account, following the argument of \cite{Khalife:2023qbu}.
Given the inapplicability of the halofit prescription for calculating non-linear matter power spectra in the late-time decaying DM models \cite{Poulin:2016nat}, we conservatively exclude the full $P(k)$ dataset from CFHTLens \cite{Heymans:2013fya} entering into the non-linear scales up to $k\sim 5$ h/Mpc.
In practice, adding the LSS data to the above data sets can place stronger constraints on the early-time effect as illustrated by the effects of galaxy clustering data on $H_0$ in the EDE \cite{Ivanov:2020ril},
which lead to smaller value of $H_0$ than the results presented below.

\begin{table}
\begin{center}
\begin{tabular}{ccccc}
\hline\hline
Model parameters~ & Priors (III) \\ \hline
$\log_{10}(M/m)$ & $[4,10]$  \\
$\log_{10}(\tau_{M}/\rm{s})$ &  $[3,10]$\\ 
$\Gamma_{m}$ [km/s/Mpc] & $[1,10]$ \\
\hline \hline
\end{tabular}
\caption{The parameter priors considered in Sec.\ref{nr}.}
\label{pp}
\end{center}
\end{table}

\subsection{Effects of \textbf{local $H_0$ prior}}
\label{nr}
As shown in Sec.\ref{model}, the model parameters are composed of $M$, $m$, $\tau_M$ and $\Gamma_m$. 
In practice, the numerical fit is sensitive to the mass ratio $M/m$, $\tau_M$ and $\Gamma_m$, rather than the mass parameter $M$.
In this sense we take a fixed value of $M=1$ GeV for simplicity and come back to this choice if necessary. 
The parameter priors considered in this subsection are shown in Table.\ref{pp}.

We run the chains using the Metropolis-Hastings algorithm,
which are considered to be convergent when the Gelman-Rubin criterion $R-1 < 0.02$ \cite{Gelman:1992zz} is satisfied.
We obtain statistics for the chains and plots with \texttt{Getdist} \cite{Lewis:2019xzd}.

\begin{figure}
\centering
\includegraphics[width=15cm, height=18cm]{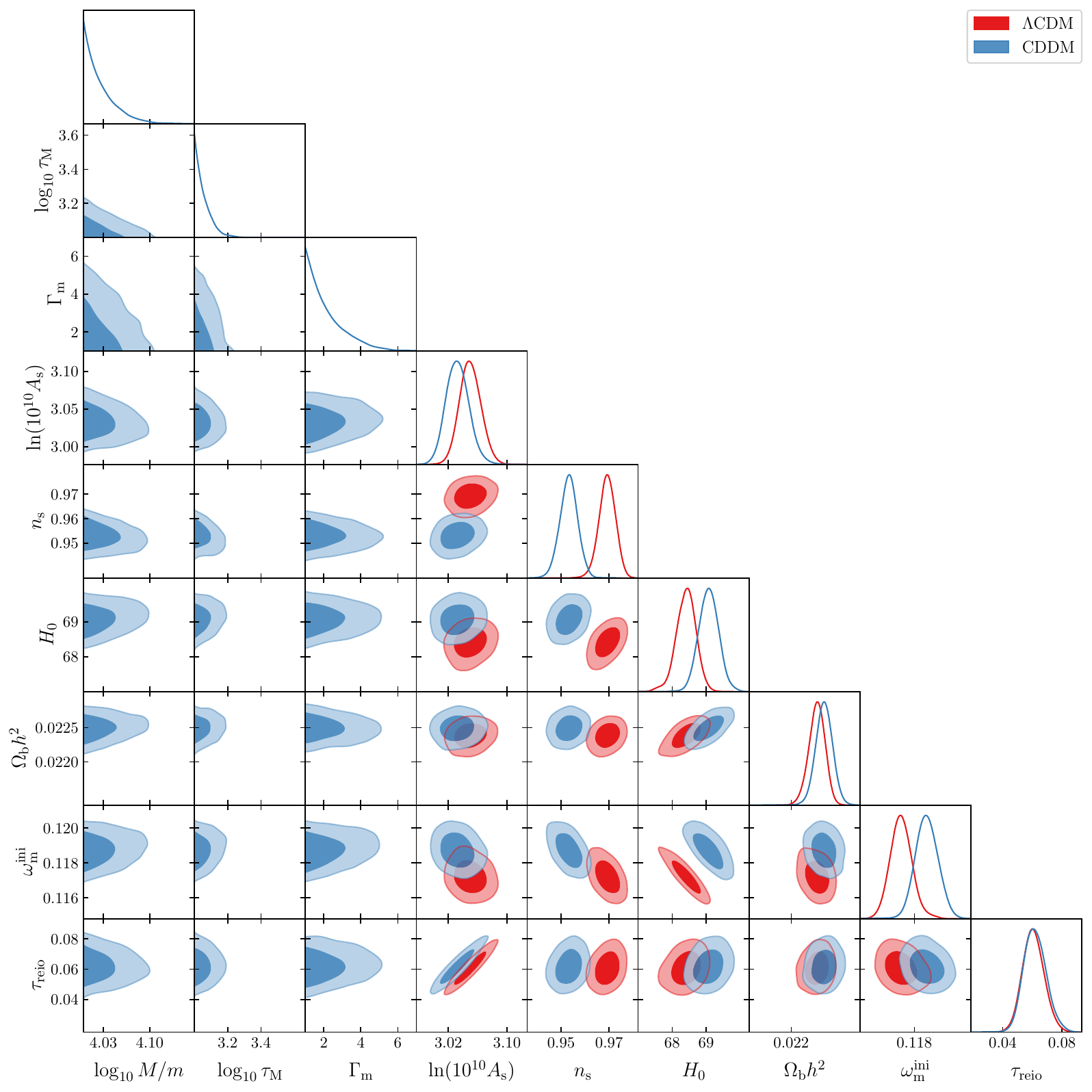}
\centering
\caption{An illustration of the MCMC fit of the CDDM model with the parameter priors (III) in Table.\ref{pp} to the datasets of Planck 2018+DESI BAO+Pantheon+SH0ES, as compared to the $\Lambda$CDM. The dark-shaded and light-shaded contours refer to  68$\%$ and 95$\%$ CL, respectively.}
\label{mcmc}
\end{figure}

Fig.\ref{mcmc} shows the complete chain of the MCMC fit of the CDDM model with the parameter priors in Table.\ref{pp} to the explicit datasets of Planck 2018+DESI BAO+Pantheon+SH0ES. 
The results  reveal two main points.
\begin{itemize}
\item First, all of the three model parameters are upper bounded, 
implying that small deviations from the $\Lambda$CDM are statistically preferred. 
We will further explore this point in the next subsection.
\item Second, Fig.\ref{mcmc} shows a 68$\%$ CL value of $H_{0}=69.05^{+0.31}_{-0.27}$  km s$^{-1}$Mpc$^{-1}$,
resulting in a reduction of the Hubble tension to $\sim 3.8\sigma$. 
\end {itemize}

 Excluding the \textbf{local $H_0$ prior}, 
 one obtains the numerical results of MCMC fit to the datasets of Planck 2018+DESI BAO+Pantheon by repeating the above process.

Table.\ref{fits} presents the mean values $\pm 1\sigma$ of the cosmological parameters subject to the datasets of Planck 2018+DESI BAO+Pantheon(+SH0ES).
Three comments are in order regarding the results in Table.\ref{fits}.
\begin{itemize}
\item  First, the effects of the \textbf{local $H_0$ prior} on both the $\Lambda$CDM and CDDM results cannot be ignored.
\item Second, while the late-time DM decay effect cannot uplift the value of $H_0$ as significantly as that of the early-time DM relativistic effect, it can effectively suppress the growth of $S_{8}$ \cite{FrancoAbellan:2020xnr,Davari:2022uwd, Enqvist:2015ara,FrancoAbellan:2021sxk}, whose value is typically larger than $\sim 0.84$  in the earlier works assuming only the early-time DM relativistic effect.
\item Third, we use the Akaike Information Criterion (AIC) to indicate the statistical preference of the CDDM relative to  the $\Lambda$CDM for the datasets considered,
which is defined as $\Delta \rm{AIC}=\Delta \chi^{2}+2(N_{\rm{CDDM}}-N_{\Lambda\rm{CDM}})$ with $N$ the number of free model parameters.
Explicitly, $\Delta \rm{AIC}=22.0(18.4)$ indicates that the fit within the CDDM model, resulting in an obvious uplift of $H_0$,
 is actually disfavored compared to the $\Lambda$CDM.
\end {itemize}

\begin{table*}
 \begin{center}
  \centering
   \renewcommand{\arraystretch}{1.5}
		\begin{tabular}{|c|c|c|}
				\hline\hline
				~ &$\Lambda$CDM & CDDM (III) \\ \hline
				
				$\rm log_{10}(M/m)$
				& ---   
				& $<4.03(4.03)$\\
				
				$\rm log_{10}(\tau_{M}/\rm{s}) $
				& --- 
				& $<3.07(3.06)$\\     
				
				$\Gamma_{m}$[km/s/Mpc]
				& ---
				& $<2.22(2.37)$\\   \hline       
				
				$H_{0}$
				& $68.09\pm0.27 (68.41^{+0.30}_{-0.27})$
				& $68.76\pm0.35 (69.05^{+0.31}_{-0.27})$\\
				
				$\omega_{b}$ 
				& $0.02230\pm0.00012(0.02238^{+0.00013}_{-0.00011})$  
				& $0.02241\pm0.00012(0.02249\pm0.00014)$\\
				
				$\omega_{\rm{m}}^{\rm{ini}}$ 
				& $0.11782\pm0.00061(0.11723\pm0.00065)$
				& $0.11945^{+0.00056}_{-0.00089}(0.11872\pm0.00068)$\\
				
				$\ln{(10^{10}A_{s})}$ 
				& $3.047\pm0.015(3.050\pm0.015)$
				& $3.028^{+0.015}_{-0.016}(3.033^{+0.014}_{-0.016})$\\
				
				$n_{s}$ 
				& $0.9673\pm0.0033(0.9692\pm0.0035)$ 
				& $0.9508^{+0.0046}_{-0.0032}(0.9532\pm0.0038)$\\
				
				$\tau_{\rm{reio}}$ 
				& $0.0586\pm0.0073(0.0607^{+0.0068}_{-0.0076})$
				& $0.0596^{+0.0067}_{-0.0080}(0.0618^{+0.0070}_{-0.0080})$\\ 
				
				$S_{\rm{8}}$ 
                & $0.8103\pm 0.0076(0.8047^{+0.0076}_{-0.0080})$
                & $0.8208^{+0.0088}_{-0.0100}(0.8136\pm0.0120)$ \\				
			        \hline
				$\Delta \rm{AIC}$ & 0.0 (0.0) & +22.0(18.4)      \\ 
				\hline \hline
			\end{tabular}

 \caption{The mean values $\pm 1\sigma$ of the cosmological parameters in the CDDM models with the parameter priors (III) in Table.\ref{pp} subject to the datasets of Planck 2018+DESI BAO+Pantheon(+SH0ES), which are compared to the $\Lambda$CDM.}
  \label{fits}
 \end{center}
\end{table*}

\subsection{Effects of parameter priors}
Having discussed the effects of the \textbf{local $H_0$ prior} on the numerical results, 
we now address the impacts of parameter priors on the results by varying the parameter priors (III) in Table.\ref{pp}.
We choose two new parameter priors with weaker two-fold effects relative to the parameter priors (III) as shown in Table.\ref{ppv}.
After repeating the MCMC fit as in Sec.\ref{nr}, we show the numerical results of fitting the model parameters to the datasets of Planck 2018+DESI BAO+Pantheon(+SH0ES) in Table.\ref{ppv2}. 

A combination of Table.\ref{fits} and Table.\ref{ppv2} clearly reveals a trend that the value of $H_{0}$ increases from $68.45^{+0.31}_{-0.26}$ km s$^{-1}$Mpc$^{-1}$ in  (I) to  $69.05^{+0.31}_{-0.27}$ km s$^{-1}$Mpc$^{-1}$ (III), 
together with the increase of $\Delta \rm{AIC}$ from $6.9$ in (I) to $18.4$  in (III).
Following this trend, the Hubble tension cannot be reduced below the $3\sigma$ level without a cost of significantly increased value of $\Delta \rm{AIC}$, revising the earlier results of $H_0 \geq 70$ km s$^{-1}$ Mpc$^{-1}$ in \cite{Blinov:2020uvz, deJesus:2022pux, daCosta:2023mow},
where the effects of parameter priors on the value of $H_0$ were not taken into account.

The numerical results in Table.\ref{fits} and Table.\ref{ppv2} are altered by at most percent level when we choose $M=1$ TeV instead of $M=1$ GeV, implying the main results on the value of $H_0$ in the CDDM model as above remain.

\begin{table*}
 \begin{center}
  \centering
   \renewcommand{\arraystretch}{1.5}
		\begin{tabular}{|c|c|c|}
				\hline\hline
				~ & Priors (I) & Priors (II)  \\ \hline
               
                                 $\rm log_{10}(M/m)$
                                 & $[1,10]$ 
				& $[3,10]$ \\
			        $\rm log_{10}(\tau_{M}/\rm{s}) $ 
				& $[1,10]$ 
				& $[3,10]$ \\     
				
				$\Gamma_{m}$[km/s/Mpc]
				& $[0,10]$ 
				& $[1,10]$ \\   
           \hline \hline
			\end{tabular}
 \caption{Two new parameter priors different from those of Table.\ref{pp}.}
  \label{ppv}
 \end{center}
\end{table*}

\begin{table*}
 \begin{center}
  \centering
   \renewcommand{\arraystretch}{1.5}
		\begin{tabular}{|c|c|c|}
				\hline\hline
				~ & CDDM (I) & CDDM (II)  \\ \hline
                                               
				$\rm log_{10}(M/m) $
				& $<2.83(2.91)$ 
				& $<3.30(3.32)$  \\
				
				$\rm log_{10}(\tau_{M}/\rm{s})$
				& $<3.09(3.04)$ 
				& $<3.57(3.68)$\\     
				
				$\Gamma_{m}$ [km/s/Mpc] 
				& $<1.76(1.77)$ 
				& $<2.33(2.53)$\\   \hline       
				
				$H_{0}$
				& $68.14\pm0.28(68.45^{+0.31}_{-0.26})$ 
				& $68.19^{+0.33}_{-0.28}(68.58^{+0.27}_{-0.32})$\\
				
				$\omega_{b}$ 
				& $0.02228\pm0.00012(0.02236\pm0.00013)$ 
				& $0.02228\pm0.00012(0.02237\pm0.00013)$ \\
				
				$\omega_{\rm{m}}^{\rm{ini}}$ 
				& $0.11799\pm0.00063(0.11740^{+0.00059}_{-0.00072})$ 
				& $0.11827^{+0.00061}_{-0.00080}(0.11761\pm0.00069)$\\
				
				$\ln{(10^{10}A_{s})}$ 
				& $3.049^{+0.014}_{-0.015}(3.053\pm0.015)$ 
				& $3.046\pm0.016(3.051\pm0.016)$\\
				
				$n_{s}$ 
				& $0.9665\pm0.0035(0.9681\pm0.0036)$ 
				& $0.9632^{+0.0053}_{-0.0033}(0.9651^{+0.0047}_{-0.0037})$\\
				
				$\tau_{\rm{reio}}$ 
				& $0.0597^{+0.0066}_{-0.0077}(0.0620^{+0.0071}_{-0.0081})$ 
				& $0.0598\pm0.0078(0.0629^{+0.0070}_{-0.0084})$ \\ 
				
				$S_{\rm{8}}$ 
                & $0.8034^{+0.0101}_{-0.0085}(0.7981^{+0.0102}_{-0.0087})$ 
                & $0.8027\pm0.0111(0.7966^{+0.0100}_{-0.0098})$  \\				
				
				\hline
				
				$\Delta \rm{AIC}$  & +6.9(6.9) & +8.0(9.7)   \\ 
				\hline \hline
			\end{tabular}
 \caption{The mean values $\pm 1\sigma$ of the cosmological parameters in the CDDM model with parameter priors (I) and (II) in Table.\ref{ppv} subject to the datasets of Planck 2018+DESI BAO+Pantheon(+SH0ES).}
  \label{ppv2}
 \end{center}
\end{table*}

\section{Complementary tests}
\label{cons}
In this section we consider complementary constraints on the parameter regions favored by the cosmological datasets.

\subsection{BBN}
\label{BBN}
An electromagnetic energy release \cite{Holtmann:1998gd, Kawasaki:2000qr} due to the decay of  $\chi_M$ during the epoch of BBN can alter the relic densities of light elements.
As a result, the measured relic densities of light elements place constraints \cite{Cyburt:2002uv, Feng:2003uy} on the $\chi_M$ decay lifetime and the electromagnetic energy release parameter $\zeta_{\rm{EM}}$ defined as
\begin{equation}\label{zeta1}
\zeta_{\rm{EM}}=\epsilon_{\rm{EM}}Y_{M},
\end{equation}
where $\epsilon_{\rm{EM}}$ is the initial electromagnetic energy released in each $\chi_M$ decay in eq.(\ref{decay1}),
and $Y_{M}=n_{M}/n_{\gamma}$ is the number density of $\chi_M$ before they decay, normalized to the number density of background photons  $n_{\gamma}$.

\begin{figure}
\centering
\includegraphics[width=12cm,height=8cm]{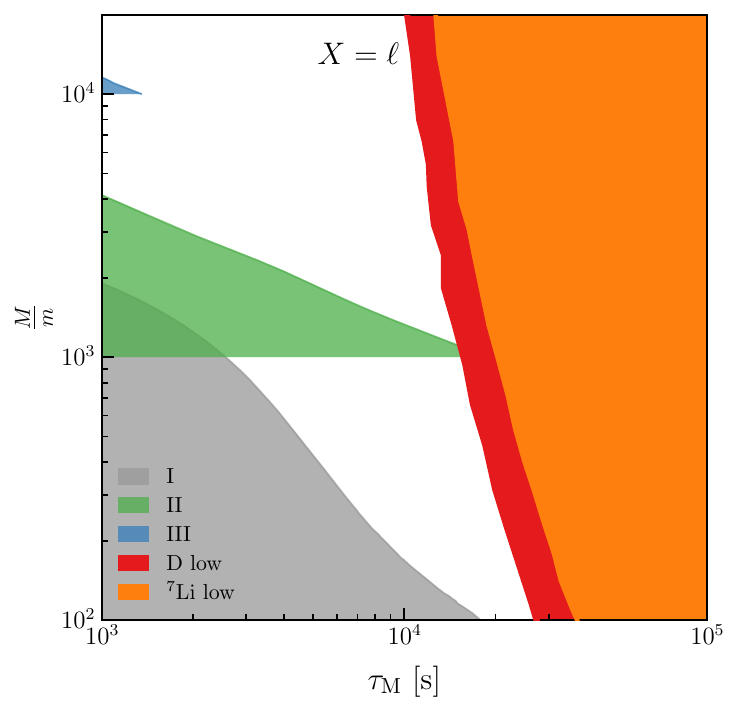} 
\centering
\caption{The 68$\%$ CL contours of $(\tau_{M}, M/m)$, extracted from the numerical results of Sec.\ref{MCMC}, compared to the BBN constraints with $X=\ell$ \cite{Cyburt:2002uv}.}
\label{bbn}
\end{figure}

In the situation with $M\gg m$ and $M\gg m_{X}$ considered here, 
$\epsilon_{\rm{EM}}\approx M/2$
for $X=\ell $ in eq.(\ref{decay1}) with $\ell$ the SM leptons\footnote{
For $X=h$ which mainly decays to the $b$ quark pair, the hadronic injection requires $M/m \leq 1$ \cite{Kawasaki:2017bqm} with $\tau_{M} \sim 100$ s, 
which is not viable for our purpose. 
On the other hand, for $X=\nu$ the BBN limits can be further relaxed, see e.g, \cite{Bianco:2025boy}.} for illustration.
Inserting the value of $\epsilon_{\rm{EM}}$ into eq.(\ref{zeta1}) gives 
\begin{equation}\label{zeta3}
\zeta_{\rm{EM}}\approx 1.5\times 10^{-9} \rm{GeV} \left(\frac{M}{m}\right),
\end{equation}
where $Y_{M}\approx \Omega_{\rm{dm}}\rho_{c}/(mn_{\gamma, 0})$ has been used.
Fig.\ref{bbn} shows the BBN constraints on the plane of $(\tau_{M}, M/m)$ for the CDDM models with parameter priors (I)-(III),
where eq.(\ref{zeta3}) has been used to transfer the BBN limits on $\zeta_{\rm{EM}}$ to $M/m$ for an explicit $\tau_{M}$.
Here, the shaded regions are excluded by $\rm{D}/\rm{H}< 1.3\times 10^{-5}$ and $^{7}\rm{Li}/\rm{H}<0.9\times 10^{-10}$ \cite{Cyburt:2002uv}.
All of the three 68$\%$ CL contours, extracted from the numerical results of Sec.\ref{MCMC}, are consistent with the BBN constraints.

\subsection{Neutrino flux}
\label{Neutrino}
\begin{figure}
\centering
\includegraphics[width=12cm,height=8cm]{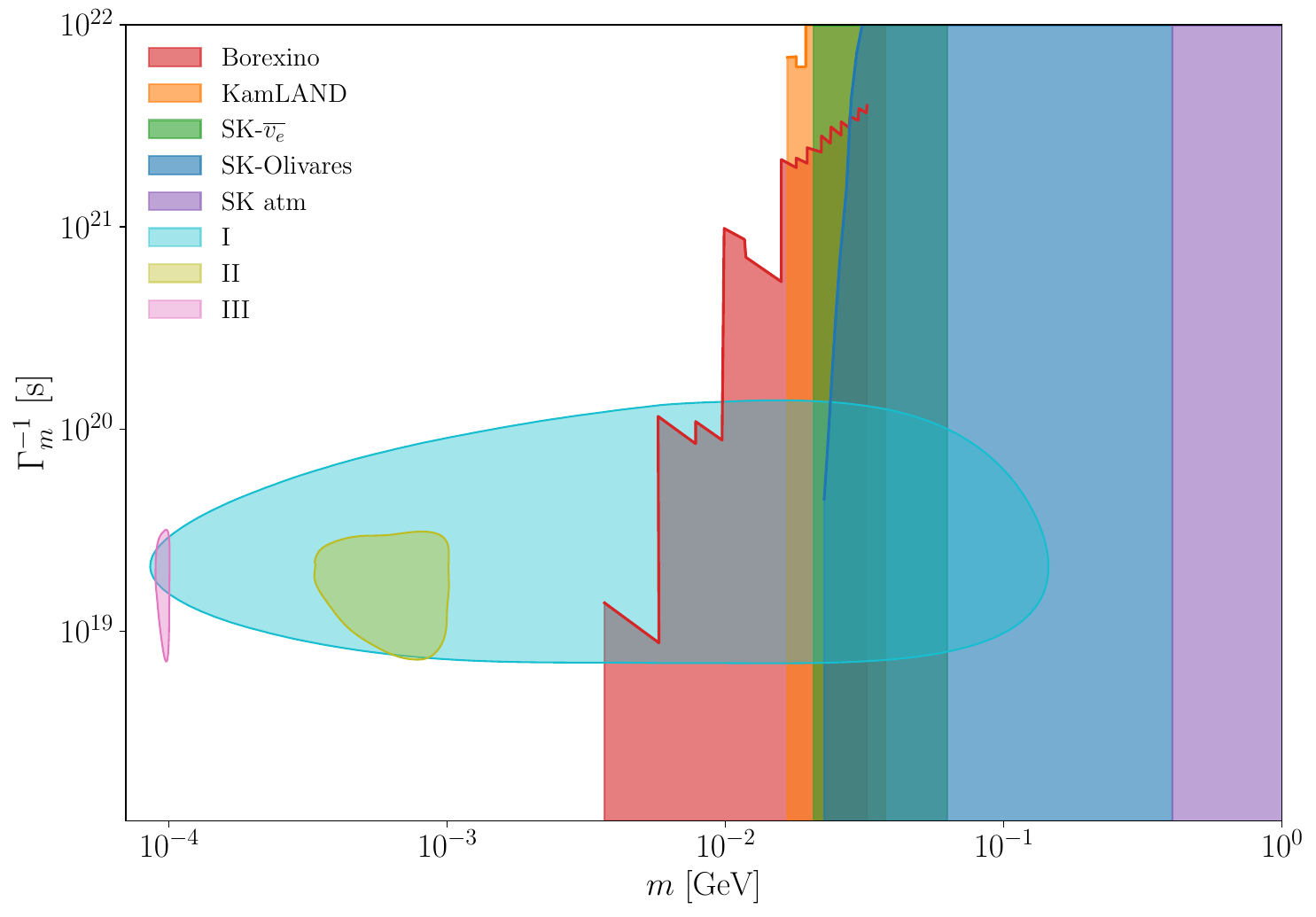}
\centering
\caption{Neutrino flux constraints \cite{Arguelles:2022nbl}  on the 68$\%$ CL contours  of $(m,\Gamma^{-1}_{m})$ arising from  Borexino \cite{Borexino:2019wln},  KamLAND \cite{KamLAND:2021gvi}, SK-$\bar{\nu}_{e}$ \cite{W. Linyan}, SK-Olivares \cite{Olivares-DelCampo:2017feq} and SK atm  \cite{Super-Kamiokande:2015qek}.}
\label{neutrino}
\end{figure}

The late-time DM decay into neutrinos contributes to neutrino flux constrained by various neutrino telescope data.
Since the numerical results in Sec.\ref{MCMC} are sensitive to $M/m$ rather than $M$,
the DM mass $m$ is therefore unfixed.
Fig.\ref{neutrino} shows the 68$\%$ CL contours  of $(m,\Gamma^{-1}_{m})$ extracted from Table.\ref{fits} for the parameter priors (III)  and  from Table.\ref{ppv2} for the parameter priors (I) and (II), respectively.  
Compared to the existing bounds \cite{Arguelles:2022nbl} including Borexino \cite{Borexino:2019wln},  KamLAND \cite{KamLAND:2021gvi}, SK-$\bar{\nu}_{e}$ \cite{W. Linyan}, SK-Olivares \cite{Olivares-DelCampo:2017feq} and SK atm  \cite{Super-Kamiokande:2015qek},
the lower bounds on $\Gamma^{-1}_m$  are converted to upper mass bounds on $m$, pointing to the mass range of $m< 5$ MeV.

\subsection{Structure formation}
\label{sf}
The $\chi_M$ decay gives rise to highly boosted DM \cite{Borah:2025cqj} at $z\geq 10^{4}$, 
which may be subject to the structure formation constraint on DM free-streaming length.
This decay induced kick velocity of DM  is given by
\begin{eqnarray}
v _{kick}=\left | v_{\chi _{m} } \right | =\frac{M^{2}-m^{2} }{M^{2}+m^{2}}.
\end{eqnarray}
For $a_{D} \ll a_{nr} < a_{\rm{eq}} $ where $a_{nr}\equiv p/m$ with $p=a p_{\rm{max}}$ and $p_{\rm{max}}$ the maximum physical momentum, the free-streaming length reads as
\begin{eqnarray}
\lambda_{fs}= \int_{a_{D}}^{a_{\rm{eq}}}\frac{da^{'}}{a^{'}}\frac{1}{a^{'}H\left(a^{'}\right)}\frac{v_{kick}\,a_{D}}{a^{'}} 
=\int_{a_{D}}^{a_{\rm{eq}}}\frac{da^{'}}{a^{'3}}\frac{v_{kick}a_{D}}{H_{0}\sqrt{\Omega_{r}a^{'-4}}}=\frac{v_{kick}}{a_{D}H(a_{D})}\ln\left ( \frac{a_{\rm{eq}}}{a_{D}} \right ).
\end{eqnarray}

Table.\ref{freestream} shows the values of $\lambda_{fs}$ using the best-fit values of cosmological parameters in the CDDM (I) to (III).
Therein the values of $\lambda_{fs}$ are of order $\sim 10^{-2}$ Mpc, being consistent with the structure formation constraint $\lambda_{fs}<0.1$ Mpc \cite{Irsic:2017ixq,DES:2020fxi,Villasenor:2022aiy}.

\begin{table}
 \begin{center}
  \begin{tabular}{ccccccccc}
   \hline\hline
   ~& &$v_{kick}$ ~&$a_{D}$ ~&$a_{nr}$ ~&$a_{eq}$ ~&$\lambda_{fs}$(Mpc)\\
   \hline
   & I   &0.999 &$5.08\times 10^{-8}$ &$8.50\times 10^{-7}$ &$2.98\times 10^{-4}$ &$4.92\times 10^{-2}$\\
   & II   &0.999 &$1.50\times 10^{-8}$ &$1.01\times 10^{-5}$ &$2.98\times 10^{-4}$ &$1.33\times 10^{-2}$\\
   & III   &0.999 &$1.66\times 10^{-8}$ &$8.39\times 10^{-5}$ &$2.96\times 10^{-4}$ &$1.22\times 10^{-2}$\\
   \hline \hline
  \end{tabular}
  \caption{The values of $\lambda_{fs}$ in the CDDM model.}
  \label{freestream}
 \end{center}
\end{table}

\section{Conclusion}
\label{con}
The Hubble tension favors an early-time and/or late-time modification to the $\Lambda$CDM.
To date, reducing this tension to $3\sigma$ level or lower has been achieved in a few scenarios such as the EDE and ad hoc combinations of the two-fold effects.
In this work, we have revisited the latter possibility in the framework of CDDM model.

We have performed a systematic analysis on the CDDM model with the following results:
\begin{itemize}
\item We parametrize the CDDM model using three effective parameters, namely $M/m$, 
$\tau_{M}$, and $\Gamma_{m}$,
through which the background and perturbation equations can be implemented into a Boltzmann solver. 
\item We carry out the MCMC fit of this model to the latest datasets, including the DESI BAO,
with an emphasis on the effects of the \textbf{Local $H_0$ prior} on the numerical results.
\item We investigate the impacts of parameter priors on the numerical results and show that 
the Hubble tension cannot be reduced below the $\sim 3\sigma$ level without a cost of significantly increased value of $\Delta\rm{AIC}$.
Our findings revise the earlier results on the tension level in the literature.
\item Finally, we discuss the complementary constraints from the BBN, neutrino flux, and structure formation,
showing that the parameter regions favored by the cosmological datasets are compatible with these existing limits.
\end{itemize}

Given the fine tuning issues of the existing solutions to the Hubble tension, 
resolving this problem remains a significant challenge.

\section*{Acknowledgements}
We thank the anonymous referee for constructive suggestions and comments. 
The authors acknowledge the use of codes \texttt{CLASS} \cite{Lesgourgues:2011re,Blas:2011rf}, \texttt{Cobaya} \cite{Torrado:2020dgo,cobaya2019} and \texttt{Getdist} \cite{Lewis:2019xzd}.

\end{document}